\documentclass[twocolumn,superscriptaddress,floatfix,showpacs]{revtex4-1}
\usepackage{amsmath,amsfonts,amssymb}
\usepackage[pdftex]{color,graphicx,hyperref}
\pdfoutput=1
\usepackage{subfigure}
\usepackage{graphicx}
\usepackage{multirow}
\usepackage{array}
\usepackage{colortbl}
\begin{document}

\title{Correlated dynamics in egocentric communication networks}
\author{M\'arton Karsai}
\email{marton.karsai@aalto.fi} 
\author{Kimmo Kaski}
\affiliation{BECS, School of Science, Aalto University, P.O. Box 12200, FI-00076}
\author{J\'anos Kert\'esz}
\affiliation{Institute of Physics, Budapest University of Technology and Economics, H-1111 Budapest}
\affiliation{BECS, School of Science, Aalto University, P.O. Box 12200, FI-00076}
\date{\today}

\begin{abstract}
We investigate the communication sequences of millions of people through two different channels and analyze the fine grained temporal structure of correlated event trains induced by single individuals. By focusing on correlations between the heterogeneous dynamics and the topology of egocentric networks we find that the bursty trains usually evolve for pairs of individuals rather than for the ego and his/her several neighbors thus burstiness is a property of the links rather than of the nodes. We compare the directional balance of calls and short messages within bursty trains to the average on the actual link and show that for the trains of voice calls the imbalance is significantly enhanced, while for short messages the balance within the trains increases. These effects can be partly traced back to the technological constrains (for short messages) and partly to the human behavioral features (voice calls). We define a model that is able to reproduce the empirical results and may help us to understand better the mechanisms driving technology mediated human communication dynamics.
\end{abstract}
\maketitle

\section*{Introduction}
\label{section:introduction}

\begin{figure}[ht!]
  \includegraphics[width=82mm]{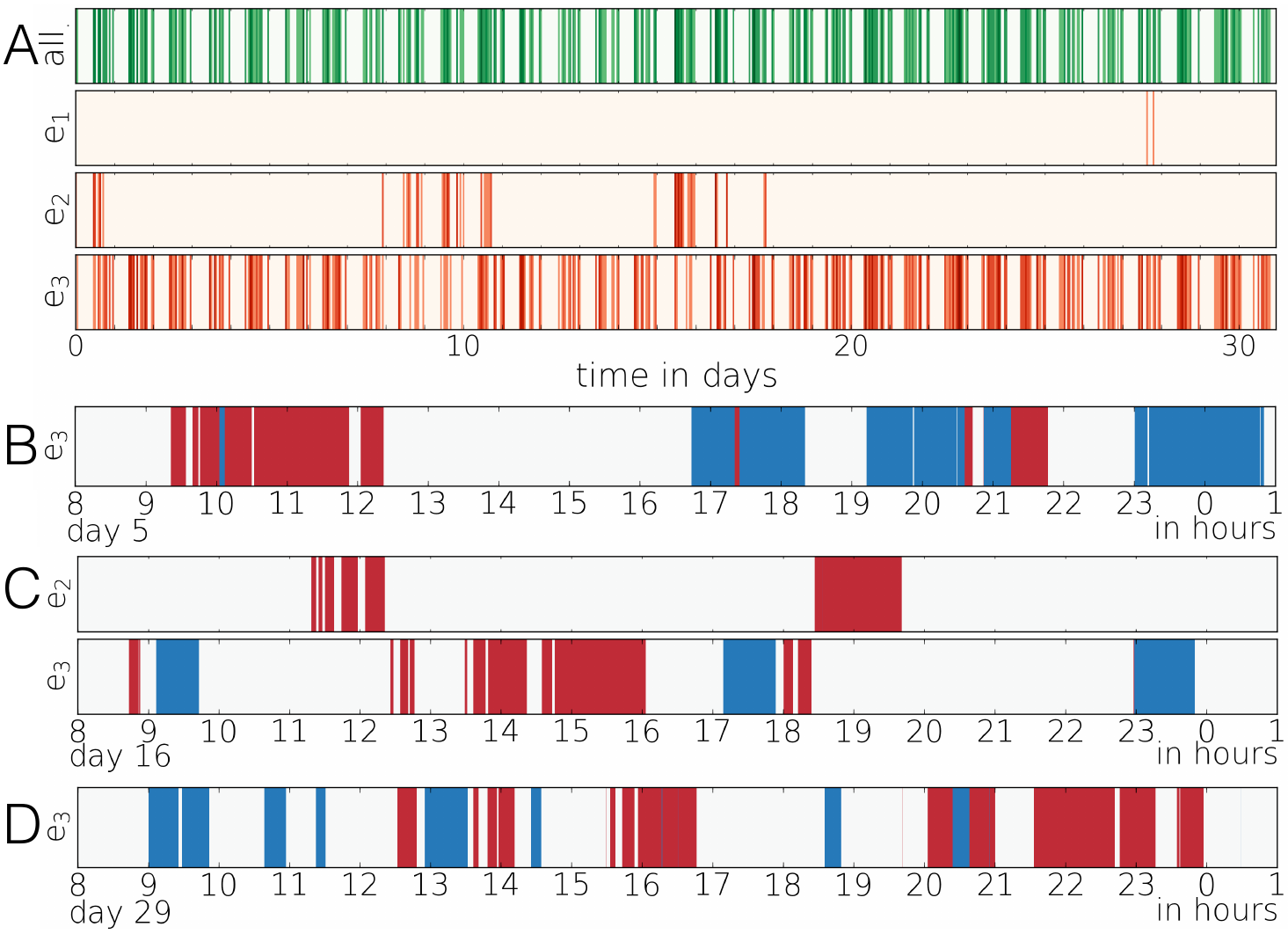}
\caption{{\bf Call activity in an egocentric network.} (a) Overall activity: darker the color larger the number of calls at the given hour. First line (green) denotes the overall in/out communication of the selected user, while the lines $e_1,e_2$ and $e_3$ demonstrate the communication on each active edges. (b,c,d) In- (red) and outgoing (blue) activity on given edges during selected time periods. The width of the coloured strips reflects the length of the corresponding calls. The selected periods are between 8AM and 1AM on the 5th, 16th and 29th days.}
\label{fig:schem}
\end{figure}

Egocentric networks consist of a central node (the ego) and its immediate neighbors. They are broadly used in psychology and sociology as they are crucial in understanding the interactions between an individual and his/her proximate social circle \cite{Wasserman1,Newman1}. Dunbar discovered that the number of social ties of an individual is anthropologically determined to be around 150 \cite{Dunbar1,Hill1,Roberts1} and that the ties can be classified into self-containing intimacy circles \cite{Barrett1} based on the strength of one's relationships. These differences in the importance of the ties of an ego are reflected by the dynamics of social relationships. An individual interacts heavily only with very few people like close kins or closest friends and communicates less with most of his/her acquaintances depending on their social closeness and common tasks \cite{Palchykov1}. This diversity is naturally present also in the communication dynamics thus data collecting communication sequences of individuals can even be used to measure the intensity of relationships \cite{Onnela1,Onnela2}. Fig.\ref{fig:schem}.a illustrates the communication pattern in an egocentric network, where the overall activity of the ego (green row) and activities on separated edges with three friends (orange rows) are presented. After building up the complete social network from egocentric subgraphs it turns out that strong links with heavy communication are usually found inside densely connected groups or communities as those people who are connected through an important relationship share many common neighbors. At the same time weak links more likely connect communities \cite{Granovetter1,Onnela1,Onnela2}, which makes them important for processes evolving in social networks \cite{Karsai2}.

Individual human behaviour shows heterogeneity not only in topology but also in dynamics. The actions of a person are not evenly distributed but rather clustered in time (e.g. see Fig.\ref{fig:schem}.a). It commonly happens that several events are executed in bursts within a short time frame and such high activity periods are separated by long inactive ones. This kind of temporal inhomogeneity can be observed in various human activity sequences ranging from communication to library loans or printing logs \cite{Oliveira1, BarabasiBursts, Goh1, Eckmann1} and they are typically characterized by the broad distribution of inter-event times $P(t_{ie})\sim t_{ie}^{-\gamma}$. In the past few years different mechanisms were proposed to describe the origin of these heterogeneities \cite{Barabasi1, Vazquez1, Malmgren1,Anteneodo1, Wu1}, and the role of external influences e.g. circadian patterns have also been addressed \cite{Jo1}.

Even though the inter-event time distribution assigns temporal inhomogeneities, it does not characterize possible correlations in the sequences. It was observed that such correlations between bursty events, indeed, exist as indicated by trains of bursty events in Fig.\ref{fig:schem}.b-d, where the incoming and outgoing calling activity of an individual are shown on each of his/her links for some selected time periods. By appropriately defining such correlated trains it turned out that the distribution of the number $E$ of events they contain shows a $P(E)\sim E^{-\beta}$ scaling \cite{Karsai1}. This behavior has been found in various human communication sequences and even in earthquake and neuron firing statistics, indicating same kind of universality behind these correlated heterogeneous dynamics. These phenomena were interpreted as a result of memory effects and modeled \cite{Karsai1} by simple multi-scale reinforcement processes similar to that describing face-to-face interactions \cite{Cattuto:2010,Stehle:2010}.

In spite of some related works \cite{Wu1} the fine grained structure of correlated trains in communication sequences are still mainly unexplored. In this paper by using detailed electronic records of large number of individuals we aim to study this problem and addressing the following questions: How are the event trains correlated with the skeleton of the backgrounding social network? Are they results of collective behavior or are they induced by individuals? Are there any correlations between the directions of the interactions? This paper is organized as follows. After introduction we define the basic quantities and show that the correlated bursty communication pattern can be attributed to links rather than to nodes. In the subsequent section we compare the balance of in- and outgoing calls and SMS-s on a link to that of the trains on the same link. We then present a model study to gain deeper insight into the governing rules of the observed behavior. After drawing the conclusions we close the paper by the detailed description of the investigated datasets and used methods.

\section*{Dynamics in egocentric networks}
\label{section:dynamics}

\begin{figure}[ht!] \centering
\includegraphics[width=82mm]{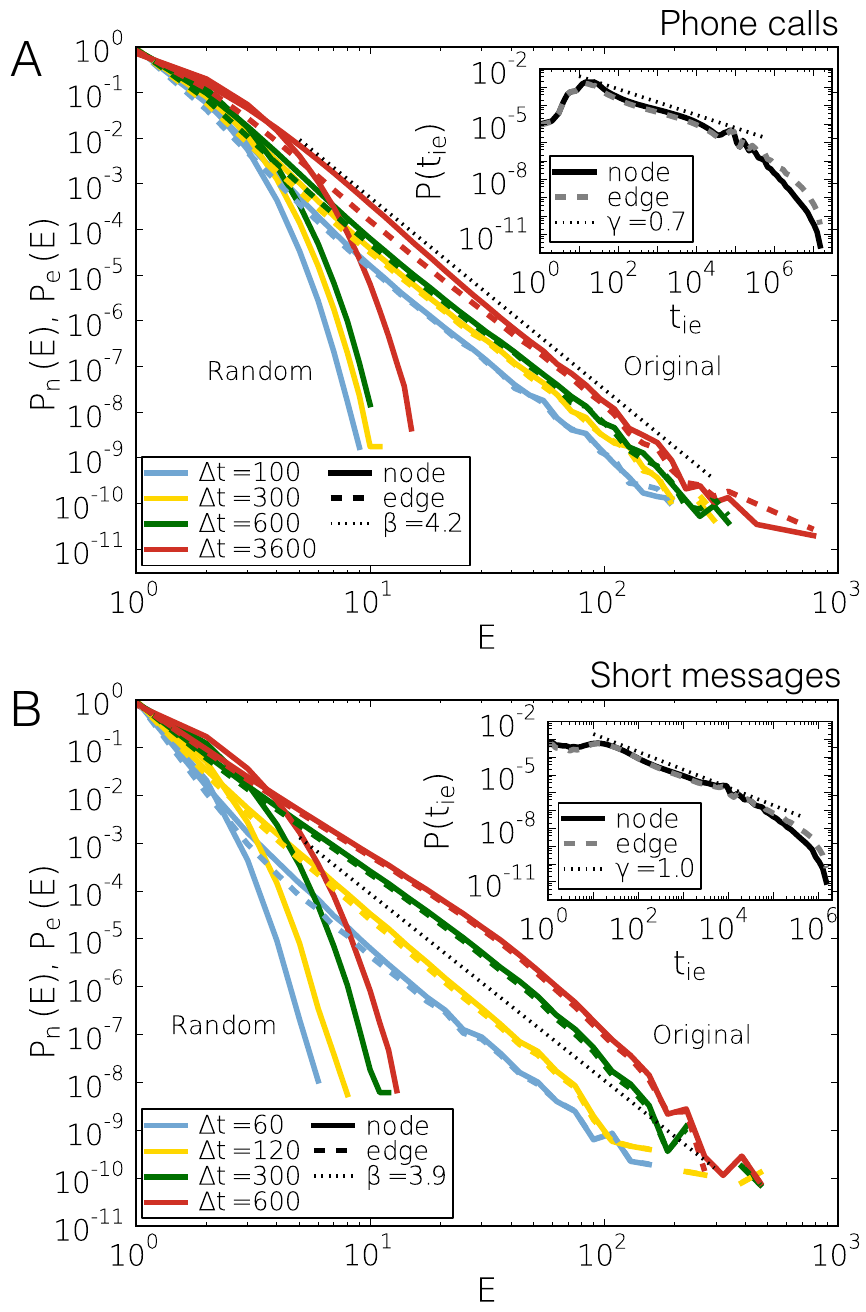}
\caption{{\bf Characteristic distributions of call and SMS sequences.} (a) Distributions of $E$ length of bursty periods of outgoing events of nodes towards all the neighbors (solid lines) and to the direction of single neighbors (dashed lines). The measurement was repeated with time windows $\Delta t=100, 300, 600$ and $3600$ seconds using the original event sequence (original) and after the inter-event times were shuffled (random). Inset: Inter-event time distributions between consecutive outgoing events of the same user towards all neighbors (solid line) and between outgoing events on a single link (dashed line). (b) The same measurements repeated for outgoing short messages with $\Delta t=60, 120, 300$ and $600$ seconds.}
\label{fig:PE1}
\end{figure}

In \cite{Karsai1} it has been demonstrated that the correlations in a bursty time series showing a power law inter-event time distribution $P(t_{ie})$ \cite{BarabasiBursts, Goh1, Eckmann1} can be detected by looking for bursty event trains and investigating the distribution $P(E)$ of the number $E$ of events they include.  A train of bursty events consists of consecutive events all with inter-event times $t_{ie}\leq \Delta t$ inside the train, separated from the rest by $t_{ie} > \Delta t$. For any independent $P(t_{ie})$ (including power law ones) $P(E)$ will decay exponentially while for correlated signals a power-law behavior of $P(E)$
\begin{equation}
P(E)\sim E^{-\beta}
\label{eq:1}
\end{equation} 
appears to indicate strong correlation of bursty events. In the present case for voice calls (SMS) both $P(t_{ie})$ and $P(E)$ are distributed like a power-law with exponents $\gamma\simeq 0.7$ ($\gamma\simeq 1.0$)  and $\beta=4.2$ ($\beta=3.9$) (see original solid lines in Fig.\ref{fig:PE1}). This behaviour is remarkably different from the $P(E)$ in sequences where inter-event times of the whole data set are randomly shuffled. This procedure leaves the inter-event time distribution and the number of communication events between pairs of users unaltered but destroys all temporal correlations. Correspondingly, the $P(E)$ distribution decays exponentially (random solid lines on Fig.\ref{fig:PE1}.a and b).

If we assume burstiness to be a mere consequence of individual human behavior, it is natural to concentrate on outgoing voice calls and SMS'. However, beyond studying only the timings of events, we can look at how they are distributed in the egocentric topology. Intuitively one may assume that correlated outgoing calls or SMS' of an individual serve the information processing of a group, thus these events are directed towards several neighbours leading to the evolution of larger temporal motifs. The existence of such behaviour was demonstrated in \cite{Kovanen2} and also in Fig.\ref{fig:schem}.c where beyond the ego two other friends are involved in some bursty sequences. Supposing this mechanism be relevant then the burstiness would be the property of a single node or a group of individuals.

Surprisingly, the generic picture seems to be different. If we assume that the correlated events in a train are directed toward several neighbours, trains of events on single edges between two persons should have an entirely different, less correlated statistics of bursts. However, this is not the case as the $P(E)$ distributions which were detected on single edges are scaling similarly and can be characterized by the same exponent values as in the case, when calls on any egocentric edges were taken into account. This suggests that trains of events usually evolve on single links. Such behaviour was confirmed both for calls and SMS' (Fig.\ref{fig:PE1}.a and b), where the $P(E)$ distribution of trains evolving between pair of individuals (original dashed lines) are only slightly different from the $P(E)$ distributions of trains which can involve several people (original solid lines). This picture is also supported by the statistics of temporal motifs \cite{Kovanen2}, where motifs involving two individuals are by far the most common ones.

\begin{figure}[ht!] \centering
\includegraphics[width=82mm]{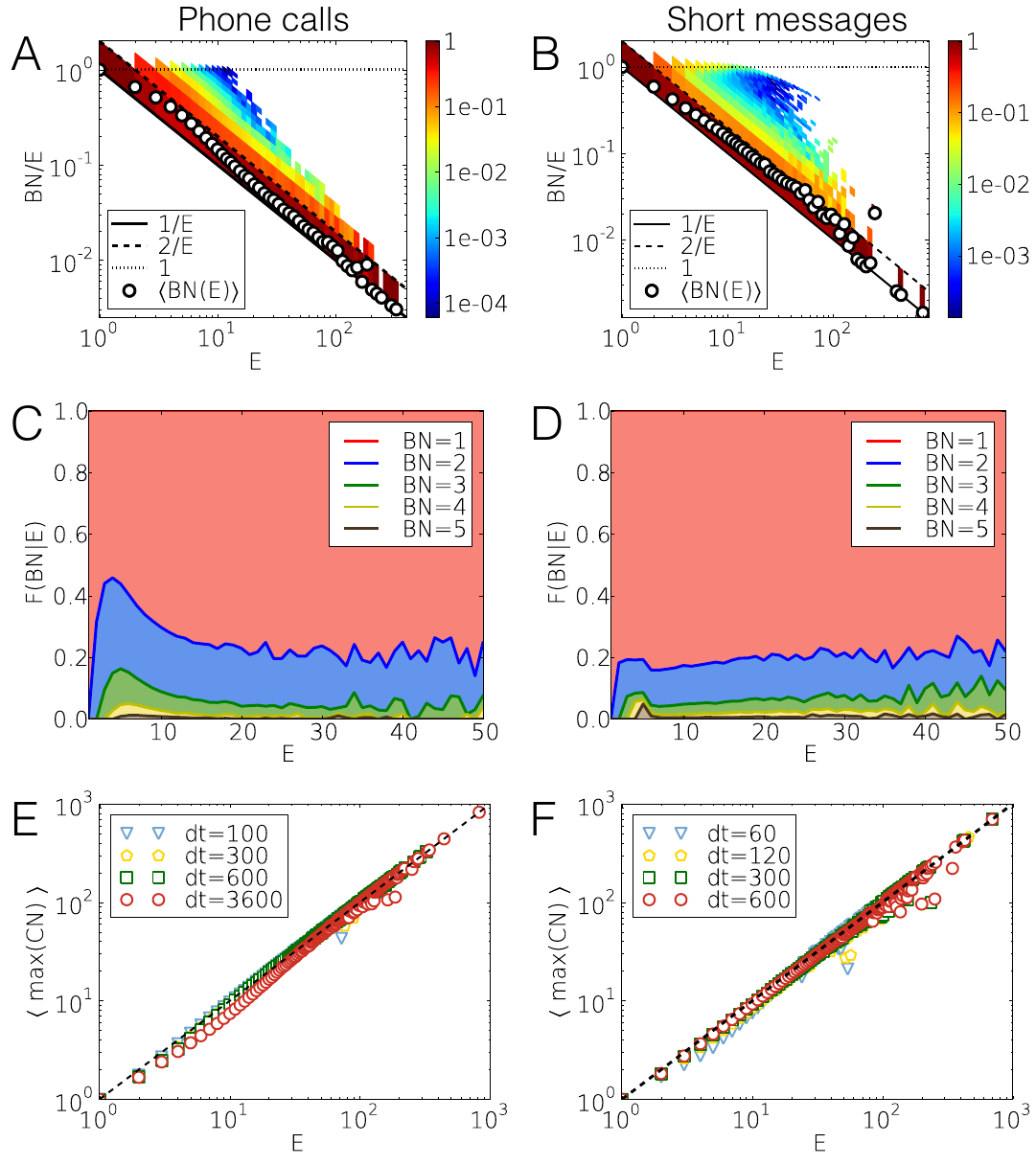}
\caption{{\bf Measures of single link bursty trains.} (a, b) The distribution and average values of the ratio $BN/E$ ($BN$ being the number of neighbours, whom an individual called in a bursty train) for each $E$ train size. Trains were detected with $\Delta t=600$ ($\Delta t=300$) for calls (SMS). The pointed and solid lines denote the limiting case $1$ and $1/E$, while dashed line belongs to $2/E$. (c, d) Cumulative fraction of the number of bursty neighbours (BN) for trains with size $E$ for calls and SMS. (e, f) Average of the maximum number of events directed to the same user  within a bursty period in case of calls and SMS.}
\label{fig:3}
\end{figure}

The same conclusion can be drawn by counting the number of neighbours $BN$, whom an individual called in a bursty train of $E$ events. If a user communicates with only one neighbour during the period then the ratio $BN/E=1/E$ or, e.g., if each calls are directed toward different neighbours than $BN/E=1$. The distributions of the $BN/E$ ratios for each $E$ together with the average values are presented in Fig.\ref{fig:3}.a and b for calls and SMS, respectively. For shorter trains the distributions show some dispersion, however the mean values confirm that usually only one or two people are called in a bursty train. To estimate what fraction of trains of the same size involves one, two or more people, we calculate the cumulative fraction of trains involving $BN$ number of neighbours. As it is depicted in Fig.\ref{fig:3}.c and d, for any $E$ the vast majority of the trains of events involve only one neighbour (red areas) while the fractions with multiple call receivers are considerably smaller. In addition Fig.\ref{fig:3}.e and f indicate that even though trains consist of events executed with several friends, most of the calls are performed with only one of them. This can be concluded as the average of the maximum number of calls performed with the same neighbour in a train with size $E$ goes with the train size as $\langle max(CN) \rangle (E)\sim E$.

\section*{Mutual bursty behaviour}
\label{section:balance}

In the previous section we pointed out similarities in the dynamics of communication between calls and SMS'. In both cases the temporal distributions of events are strongly heterogeneous, the sequences consist of correlated actions which are clustered into long bursty trains, and which trains are usually evolving on single links rather then connecting a larger group of people. Now we are at the position to ask further questions about the dynamics of human interactions on a dyadic level like: What is the direction of events in the trains? Is the communication in bursty periods balanced or dominated by one partner? Are there differences between voice calls and SMS' from this point of view?

\begin{figure}[ht!]
\includegraphics[width=82mm]{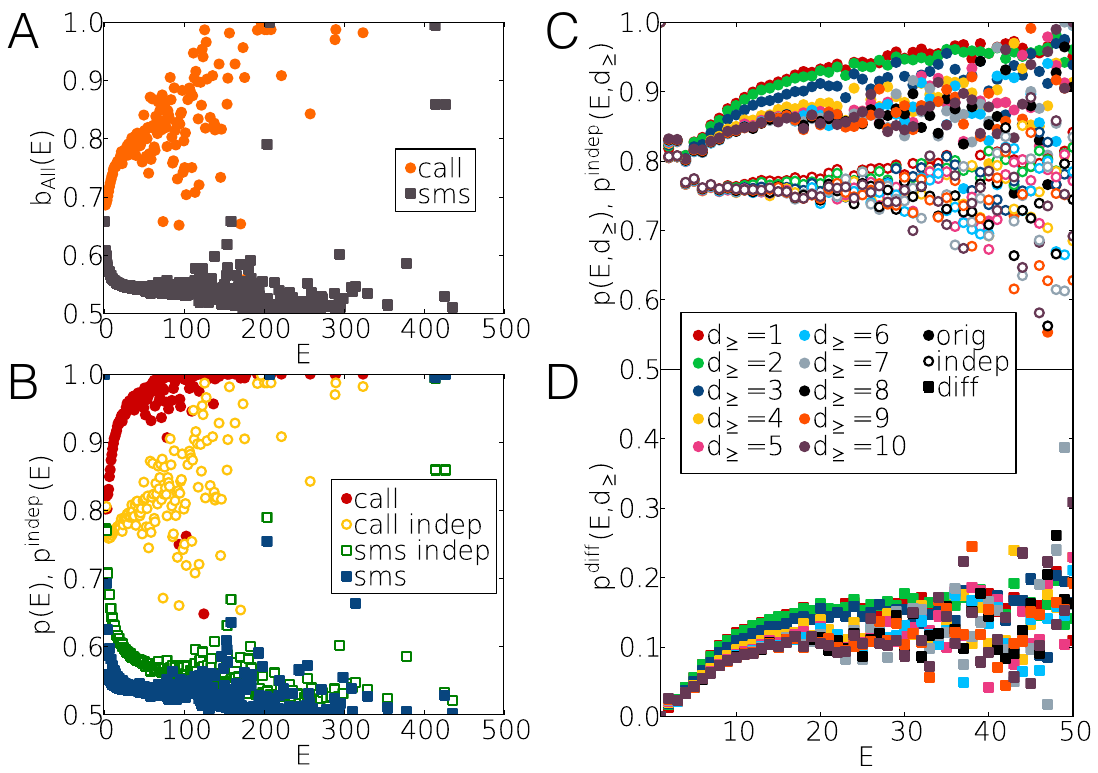}
\caption{{\bf Directed balance measures in call and SMS sequences.} (a) Average $b_e$ edge balance values calculated for trains with the same size for calls (orange circles) and SMS (brown squares). (b) Average $p(E)$ train balance values for trains of the same size in case of calls (red circles) and SMS (blue squares). A similar average was calculated for corresponding independent event trains (yellow circles for calls and green squares for SMS). (c) Average $p(E)$ and $p^{indep}(E)$ measured in call sequences where events with duration smaller than $d_{\geq}$ were removed. (d) The difference of the empirical and independent values in call sequences where events with duration smaller than $d_{\geq}$ were removed.
}
\label{fig:6}
\end{figure}

For the entire recording period of the dataset one can calculate the overall communication balance for each edge $e$ as:
\begin{equation}
\hspace{.2in} b_e=\dfrac{max(N_A,N_B)}{N_A + N_B}
\label{eq:b}
\end{equation}
where $N_A$ ($N_B$) assigns the total number of calls from $A$ to $B$ ($B$ to $A$). Hence $b_e$ can vary between $1/2$ (completely balanced) and $1$ (completely imbalanced, dominated by one of the participants). Strong difference occurs between voice call and SMS dynamics if we look at the balance of edges and their correlation with the length of trains which evolve on them. To do so, we calculate the  weighted average of $b_e$ over all trains of length $E$:
\begin{equation}
b_{All}(E)=\langle b_e\rangle_{E\ {\it trains}}=\dfrac{\sum_e n_e(E)b_e}{\sum_e n_e(E)}
\label{eq:b_All}
\end{equation}
where $n_e(E)$ means the number of trains of length $E$ on edge $e$. In Fig.\ref{fig:6}.a for SMS (brown squares) $b_{All}$ is converging to 1/2 indicating that the trains evolve on strongly balanced links and this effect is enhancing for longer trains. The balanced communication in SMS sequences has been observed by Wu. et al. \cite{Wu1} and can be explained by the technologically determined way of communication exchange, which requires mutuality for SMS. However, this constrain does not apply for the mobile calls (orange points in Fig.\ref{fig:6}.a) where during a call information can flow in both directions. Here $b_{All}$ reflects strongly unbalanced communication and it increases towards $1$ for trains with larger $E$. It assigns an opposite trend for calls compared to SMS communication as longer trains evolve on more unbalanced edges.

The next question we should address is whether the balance or imbalance within the trains simply reflect the bias due to $b_e$ of the actual edge or additional effects play role in there. To study this we define the balance within a train in a similar way as for an edge:
\begin{equation}
p_{e}(E_m)=\dfrac{max(n_A,n_B)}{n_A + n_B}.
\label{eq:p}
\end{equation}
Here $p_{e}(E_m)$ is the balance of the $m-$th train of length $E_m$ on edge $e$ connecting $A$ and $B$, $n_A$ ($n_B$) denotes the number of events initiated by $A$ ($B$) towards $B$ ($A$) in that train; $E_m=n_A+n_B$. Averaging over trains of the same size $p(E)=\langle p_e(E_m)\rangle_{E {\it \ trains}}$ gives an estimate for the average communication balance in trains of size $E$ (note that in this case  different $p_e(E_m)$ values can evolve even for trains on the same edge $e$). This has to be compared to the case, when a train is composed from events selected in an independent manner from a set with balance $b_e$. The latter can be calculated as
\begin{equation}
p_e^{indep}(E_m=E)=\dfrac{1}{E}\sum_{i=0}^{E}\left( \left| \frac{E}{2} - i \right|+\frac{E}{2} \right)b_e^i(1-b_e)^{E-i}\binom{E}{i},
\label{eq:prand}
\end{equation}
where the first factor after the summation weighting the binomial distribution %as 
taking into account that the imbalance can evolve in both directions, i.e., parallel or antiparallel to the imbalance of the edge. As $p_e^{indep}(E_m=E)$ depends only on $b_e$ and $E$, the average can be taken similarly to $p(E)$  as $p^{indep}(E)=\langle p^{indep}_e(E_m) \rangle_{E{\it \ trains}}$ to get the estimate for the independent case.

Fig. \ref{fig:6}.b shows $p(E)$ and $p^{indep}(E)$  for both voice calls and SMS messages. The first apparent feature is that for large values of $E$ the $b_{All}(E)$ of Fig.\ref{fig:6}.a and $p^{indep}(E)$ become very similar, corresponding to the fact that for long trains $p^{indep}(E)$ approaches the bias of the actual edge. However, the interesting effect is the difference between $p(E)$ and the corresponding $p^{indep}(E)$. It shows that trains of calls (red points) are much more unbalanced than one would expect from independent processes (yellow circles) considering the bias of the edge. At the same time for SMS the contrary is true as trains (blue squares) are much more balanced than %then 
one would derive from random processes (green squares) applying the $b_e$ values of the corresponding edges. This demonstrates real correlation between events of the same train and suggests different correlated mechanisms behind call and SMS dynamics. One may suspect that the strengthening of the imbalance for voice calls may be a consequence of call trials picked up by the answering machine. To exclude this %case, 
we repeated the calculations after removing events with duration shorter than $d_{\geq}$ as depicted in Fig.\ref{fig:6}.c. The calculated $p^{diff}(E)=p(E)-p^{indep}(E)$ in Fig.\ref{fig:6}.d shows that the saturation value is slightly decreasing as we remove longer calls, however, the differences remain significant and saturating to $0.1$ with $d_{\geq}> 4$ seconds. This duration time also serves as a timescale after which the answering machine calls do not play any significant role. For further details about the effect of short calls see Data and Methods.

\section*{Model}
\label{section:model}

In this section we introduce a model, which is able to reproduce the empirical observations, namely the enhanced imbalance of the bursty trains for voice calls and the opposite behaviour for the SMS message compared to independent processes. Our aim is to identify the different mechanisms controlling the dynamics of communication through these channels, integrate them into a single model which we can test against the empirical results.

The emergence of correlated bursty trains in communication sequences were interpreted earlier by memory processes \cite{Karsai1}. The evolution of trains with size scaling as Eq.\ref{eq:1} were explained as the result of reinforcement dynamics where the probability $q(n)$ to perform one more event in a train after $n$ events have been executed, depends on $n$ as:
\begin{equation}
q(n)=\left( \dfrac{n}{n+1} \right)^{\nu} \mbox{\hspace{.1in} where \hspace{.1in}} \nu=\beta+1.
\label{eq:qn}
\end{equation}
Assuming this dynamics and by fitting $\beta$ to empirical data, we are able to generate model trains with realistic size scaling. However, the question remains how we can introduce the mechanisms responsible for the enhancement of bursty trains. 

Let us first concentrate on voice calls. One correlation we observed (see Fig.\ref{fig:6}.a lower panel) is that longer trains tend to be more unbalanced, meaning that they are more dominated by one of the callers. We also disclosed the possibility (in Fig.\ref{fig:6}.b) that this behaviour is an artifact due to answering-machine calls where the caller repetitively recalls the friend to pass an important information. Keeping in mind that mobile calls enable bidirectional information change, we assume that the observed unbalanced communication in call trains reflects the difference in motivation between the communicating partners. If there is a task to solve, which is more important for one party, it gives motivation for him/her to repeated calls until the issue gets settled. 

\begin{figure}[ht!]
\includegraphics[width=82mm]{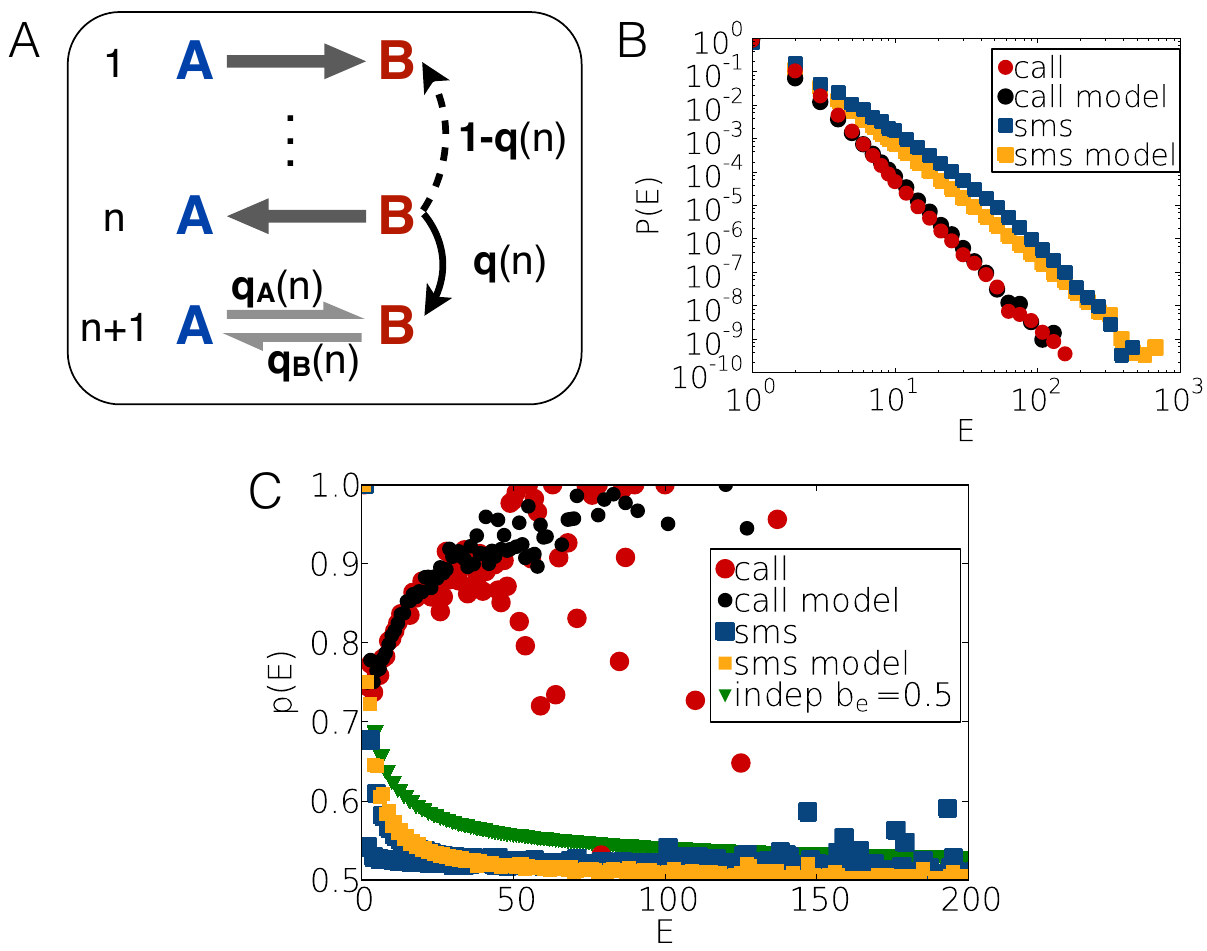}
\caption{{\bf Model definition and results.} (a) Illustrative definition of the model. Here events are simulated between two people $A$ and $B$. The dynamics and direction of the events are determined by probabilities $q(n)$, $q_A(n)$ and $q_B(n)$ defined in the text. (b) $P(E)$ distributions of empirical call trains (red circles) with $\Delta t=600s$ on edges with $0.5 \leq b_e < 0.55$ and in corresponding model trains (black circles). The same functions are shown for SMS trains (blue and yellow squares accordingly). (c) Balance values calculated for empirical call (red circles) and SMS (blue squares) trains and in corresponding model processes (red and yellow symbols). Balance values of independent trains are also shown (green triangles) calculated by Eq.\ref{eq:prand} with $b_e=1/2$.}
\label{fig:8}
\end{figure}

This mechanisms can be incorporated into the reinforcement process of bursty trains in the following way (see Fig.\ref{fig:8}.a). We simulate bursty trains, which evolve on a link between a pair of individuals $A$ and $B$. To initiate a train with a probability equal to $b_e$ we randomly select $A$ or $B$ who then perform one event towards the other agent and at the same time we set the actual train size to $n=1$. The decision about the next event is carried out in two steps. First we decide with the probability in Eq.\ref{eq:qn} whether to perform one more event in the train or initiate a new train otherwise. If the train should be continued a new decision is made about the direction of the call. The probabilities that it is initiated by $A$ or $B$ are given as:
\begin{equation}
q_{\sigma}(n|\sigma_1)=\dfrac{n}{n+1} \mbox{\hspace{.2in} or \hspace{.2in}} q_{\sigma}(n|\neg\sigma_1)=1-\dfrac{n}{n+1}
\label{eq:qAcall}
\end{equation}
where $\sigma\in \{A,B\}$. Here $q_{\sigma}(n|\sigma_1)$ denotes the probability that the $nth$ event of the actual train is performed by the same user who initiated the train at $n=1$, while $q_{\sigma}(n|\neg\sigma_1)$ gives the probability that the other player makes the call. Consequently, the longer a train evolves, the larger is the probability that the agent, who initiated the actual train will make a call to the other agent. Eq.\ref{eq:qn} and Eq.\ref{eq:qAcall} capture the coupled reasons for the evolution of long unbalanced trains. They reflect the reinforced motivation of an individual induced by the effort what he/she already invested in the actual series of calls to successfully solve a task with the other partner.

In case of SMS the mechanism for developing strong balance in bursty trains is different. There, in single events information can pass only
one way and consecutive events in a train usually have reversed direction. Trains are possibly conversations between partners and the technology constrain is responsible for the strongly balanced communication. To simulate this behaviour again we generate trains as in the previous case by using Eq.\ref{eq:qn}, however, to select the direction of the actual event we define a different mechanism. Here we assume that the direction of an event depends only on the direction of the previous one as they tend to be reversed. Accordingly, conditional probabilities can be used to decide the direction of the actual event as:
\begin{equation}
q_{\sigma}(n|\neg\sigma_{n-1})=\dfrac{n}{n+1} \mbox{, \hspace{.1in}} q_{\sigma}(n|\sigma_{n-1})=1-\dfrac{n}{n+1}
\end{equation}
where $\sigma \in \{ A,B \}$ and $q_{\sigma}(n|\neg\sigma_{n-1})$ denotes the probability to choose the opposite direction for the $nth$ event compared to the one in the $n-1th$ step. Accordingly $q_{\sigma}(n|\sigma_{n-1})$ denotes the probability of choosing the same direction as for the previous event. In this way the longer a train evolves, the larger is the probability to revert the direction of consecutive events and consequently the more the train becomes balanced.

The evolution of enhanced balance/inbalance in trains can be checked in the best way on edges where the overall communication is completely balanced and the balance/unbalance of trains is induced only by actual behavioural differences. In this way by setting $b_e=1/2$ the results obtained from the model process can be compared to averages calculated for real trains that evolve on edges with the same $b_e$ values. We selected edges with overall balance values between $0.5 \leq b_e < 0.55$ and calculate the corresponding $P(E)$ and $p(E)$ functions for this limited number of 115,277,534 calls and 69,288,504 SMS. As it is shown in Fig.\ref{fig:8}.b, the size of call trains (red circles) detected with $\Delta t=600 s$ and SMS trains (blue squares) with $\Delta t=300 s$ are distributed broadly and were characterized by an exponent $\beta=4.6$ and $\beta=3.5$ accordingly. Also the $p(E)$ balance values calculated for the limited event sets in Fig.\ref{fig:8}.c show similar behaviour to what was observed earlier in Fig.\ref{fig:6}.b for averages calculated for all edges. This figure demonstrates that even if the overall communication between people is balanced yet the strong communication unbalance for voice calls and an enhanced balance for SMS trains evolves. This is even more conspicous when we compare these curves to the values obtained for independent processes (green triangles in Fig.\ref{fig:8}.c) using Eq.\ref{eq:prand} with $b_e=1/2$.

Using the above parameters we executed model processes for calls and SMS with the same number of events and corresponding $\nu$ exponents deduced from $\beta$ according to Eq.\ref{eq:qn}. The $P(E)$ distributions calculated for the evolving model trains are fitting well to the corresponding empirical functions as it is depicted in Fig.\ref{fig:8}.b for model call trains (black circles) and for model SMS sequences (yellow squares). It should be emphasized that there is no further fitting parameter in the model, nevertheless, the average $p(E)$ balance values calculated for trains in the model processes are in very good agreement with the empirical observations in Fig.\ref{fig:8}.c. This indicates that the assumed mechanisms are capturing rather accurately the salient features of the dynamics of directed human communication through phone calls and SMS.  The only discrepancy is for the $p(E)$ values of short SMS trains, where the empirical data show an even-odd effect, which is not present in the model, indicating that for such communications an additional mechanism may be present enhancing the tendency towards the balance. 

\section*{Conclusion}
\label{section:conclusion}

In this paper we have studied mobile phone call and SMS communication sequences of millions of individuals and found the signature of internal correlations, which evolve between events and are responsible for long correlated event trains. By considering egocentric networks we realized that these bursty trains are likely to evolve between pairs of individuals rather than characterizing the communication of an individual in a larger group. Moreover, after a careful analysis we have found that the communication in such trains is much more balanced (unbalanced) for SMS (voice calls) than expected from trains of independent events with the same balance value $b_e$ of the edge, where the train was formed. We have concluded that the backgrounding mechanisms for the evolution of bursty trains can be interpreted as memory processes and implemented by reinforcement model dynamics. We have shown that events in trains are determined by different backgrounding motivations for calls and SMS; clearly in the latter case they are strongly mediated by technological constrains. We integrated these mechanisms into a model to simulate call and SMS train sequences. The model is capable in reproducing the observed statistics of events in trains of bursts and create correlations to enhance balance (imbalance) for SMS  (voice call) cascades. We believe that this work gives insight into the communication dynamics of individuals, dyads, and egocentric networks at the level of single events. It points out similarities and differences of communication through different channels and contributes to our understanding of human behaviour. 

\section*{Data and methods}
\label{section:data_methods}

In the present study we investigate the dynamics of human communication by analysing sequences of mobile-phone calls (MPC) and short messages (SMS) induced by a large set of individuals. The datasets were recorded by a single operator with $20\%$ market share in an undisclosed European country. They contain 633,986,311 time stamped MPC (209,316,760 SMS) events recorded during 182 days with 1 second resolution between 6,243,322 (4,819,993) individuals who are connected via 16,783,865 (10,339,274) edges. In order to take into account only true social interactions and avoid commercial communication, we used only actions which were executed on links between users, who are at least once mutually connected each other during the recorded period. Note that due to the partial coverage we cannot know the complete egocentric network of each individual. This may play an effect on our results, however if this impact would be significant it would be already visible from the results of the limited dataset.
\begin{figure}[h!]
\includegraphics[width=82mm]{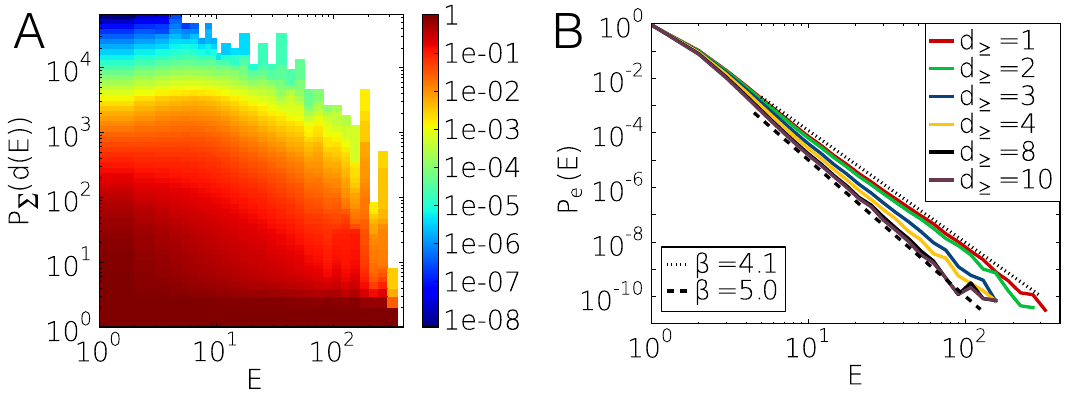}
\caption{{\bf Effect of calls with short duration.} (a) Cumulative distribution of duration of calls in bursty trains with length $E$. (b) $P(E)$ distributions of periods including events on single edges of any direction. Distributions were calculated for sequences where calls with duration smaller than $d_{\geq}$ were removed. The two straight line are denoting power-law functions with exponent $\beta=4.1$ and $5.0$.
}
\label{fig:5}
\end{figure}

To characterize temporal behaviour of an individual, we commonly used inter-event times defined as $t_{ie}=t_{i+1}-(t_{i}+d_i)$ where $t_i$ and $t_{i+1}$ denote the starting time of two consecutive outgoing call events, while $d_i$ is the corresponding duration. Naturally, in case of outgoing SMS the same definition holds but with $d_i=0$.

To filter the artificial effect of technology those consecutive short messages, which were sent to the same neighbour with inter-event times smaller than $10$ seconds were replaced with a single event as they were possibly parts of the same multipart SMS \cite{Kovanen1}. A technology related problem for voice calls could be that long trains are formed, when the caller can reach an answering machine only instead of the callee and repeats the call several times resulting in a series of rather short calls. This effect is observable in Fig.\ref{fig:5}.a, where the cumulative distribution $P_{\sum}(d(E))$ of call duration for calls detected in trains with size $E$ is shown indicating that the length of calls in longer trains are shorter. In order to check the effect of such calls on the statistics the distributions were calculated after calls with duration $d \geq d_{\geq}$ were removed (Fig.\ref{fig:5}). Altough the exponent changed from $\beta=4.1$ to $5.0$ the power-law scaling of $P(E)$ remained indicating strong residual temporal correlations.

\begin{acknowledgments}
We thank M. Kivel\"a and J. Saram\"aki for comments and useful discussions and for A.-L. Barab\'asi for the dataset used in this research. Financial support from EU’s FP7 FET-Open to ICTeCollective Project No. 238597 and TEKES (FiDiPro) are acknowledged.
\end{acknowledgments}

\textbf{Atuhors contribution}
Conceived and designed the experiments: MK, KK, JK. Performed the experiments: MK. Constructed the model: MK. Analyzed the data: MK, KK, JK. Wrote the paper: MK, KK, JK.

\end{document}